\documentclass[12pt]{article}
\begin{document}
\vbadness = 100000
\hbadness = 100000
\newcommand{\grad}{\mbox{\boldmath$\nabla$}}
\newcommand{\bdiv}{\mbox{\boldmath$\nabla\cdot$}}
\newcommand{\curl}{\mbox{\boldmath$\nabla\times$}}
\newcommand{\bcdot}{\mbox{\boldmath$\cdot$}}
\newcommand{\btimes}{\mbox{\boldmath$\times$}}
\newcommand{\btau}{\mbox{\boldmath$\tau$}}
\newcommand{\btheta}{\mbox{\boldmath$\theta$}}
\newcommand{\bphi}{\mbox{\boldmath$\phi$}}
\newcommand{\bmu}{\mbox{\boldmath$\mu$}}
\newcommand{\bepsilon}{\mbox{\boldmath$\epsilon$}}
\newcommand{\bcj}{\mbox{\boldmath$\cal J$}}
\newcommand{\bcf}{\mbox{\boldmath$\cal F$}}
\newcommand{\bbeta}{\mbox{\boldmath$\beta$}}
\newcommand{\bco}{\mbox{\boldmath$\omega$}}

 \title{Legendre polynomial expansion of the potential of a uniformly charged disk}
 \author{Jerrold Franklin\footnote{Internet address: Jerry.F@TEMPLE.EDU}\\
 Department of Physics\\ Temple University, Philadelphia, PA 19122}
  \date{\today}
   \maketitle
\begin{abstract}
We use a Legendre polynomial expansion to find the electrostatic potential of a uniformly charged disk.
We then use the potential to find the electric field of the disk.
 \end{abstract}

\section{Introduction}

Many papers have been written to derive the electrostatic potential of a uniformly 
charged disk
%\cite{1,2,3,4,5,6,7,8,9,10,11}. 
\cite{1}-\cite{11}.
However, all of these
solution methods are quite complex.
The most recent paper\cite{11} uses a Green's function method
leading to elliptic integrals, and claims that that
simplifies the problem, but it too is quite complicated.

In this paper, we use a Legendre polynomial expansion to find the potential.
This method is considerably simpler than previous derivations, and is directly 
related to Legendre polynomial expansions that graduate students would have seen in class.
The Legendre polynomial expansion for the disk can then be differentiated to find 
 the electric field of the disk.

\section{Electrostatic potential  of a uniformly charged ring }

We first derive the potential on the axis (chosen as the z-axis) of a uniformly charged ring  of charge $Q$ and radius $R$,
\begin{equation}
\phi(z)=\frac{Q}{\sqrt{z^2+R^2}}.
\end{equation}
This follows because every point on the ring is the same distance, $\sqrt{z^2+R^2}$,
from the point $z$ on the axis.

We can write the potential in terms of the variables $r$ and $\theta$ as
\begin{equation}
\phi(r,0^\circ)=\frac{Q}{\sqrt{r^2+R^2}}.
\end{equation}
This can be expanded in a binomial expansion for $r>R$ as
\begin{eqnarray}
\phi(r>R,0^\circ)&=&Q[r^2+R^2]^{-1/2}
=\frac{Q}{r}\left[1+\frac{R^2}{r^2}\right]^{-1/2}\nonumber\\
&=&Q\sum_{n=0}^\infty\left(\begin{array}{c} -1/2\\ n
\end{array}\right)\frac{R^{2n}}{r^{2n+1}}. yg7y7
\end{eqnarray}
The binomial coefficients in the expansion are defined by
\begin{equation}
\left(\begin{array}{c} a\\ n\end{array}\right)
 = \frac{a!}{n!(a-n)}.
\end{equation}

To facilitate an expansion in Legendre polynomials,
we introduce the index $l=2n$.  Then $l$ must be even and $n=l/2$,
and the expansion for $\phi(r>R,0^\circ)$ can be written as
\begin{equation}
\phi(r>R,0^\circ)
=Q\sum_{{\rm even}\; l}\left(\begin{array}{c} -1/2\\ l/2
\end{array}\right)\frac{R^{l}}{r^{l+1}}.
\end{equation}

The potential of the ring satisfies Laplace's equation for the region $r>R$.
To get the potential for all angles, we use the
 fact that the angular dependence of the $1/r^{l+1}$
  term in the expansion must be $P_l \cos(\theta)$. 
That means each term in the series that goes like $\frac{1}{r^{l+1}}$ can be extended off the axis
by simply multiplying that term by the Legendre polynomial $P_l(\cos\theta)$.
The potential at all angles can thus be given by multiplying each term in the expansion by $P_l(\cos\theta)$, so
\begin{equation}
\phi(r>R,\theta)
=Q\sum_{{\rm even}\; l}\left(\begin{array}{c} -1/2\\ l/2
\end{array}\right)\frac{R^{l}P_l(\cos\theta)}{r^{l+1}}.
\label{eq:rgR}
\end{equation}

For $r<R$ the potential still satifies Laplace's equation because there is no charge inside the ring.  We can expand
 $\phi(r<R,0^\circ)$ in powers of
$r^2/R^2$, and following steps similar to those above leads to
\begin{equation}
\phi(r<R,\theta)
=Q\sum_{{\rm even}\; l}\left(\begin{array}{c} -1/2\\ l/2
\end{array}\right)\frac{r^{l}P_l(\cos\theta)}{R^{l+1}}.
\end{equation}

Equations (6) and (7) give the  electrostatic potential of the uniformly charged ring for $r>R$ and $r<R$, respectively.

\section{Electrostatic potential  of a uniformly charged disk }

A uniformly charged disk can be considered to be composed of uniformly charged rings. The potential of rings can be integrated to find the potential of the disc.  Each ring has a radius $x$ and a charge 
$dq= [Q/(\pi R^2)]2\pi xdx$, so the potential on the axis of the disk is]\begin{eqnarray}
\phi_{\rm disk}(z)&=&\frac{2Q}{R^2}\int_0^R\frac{xdx}{{\sqrt{z^2+x^2}}}
=\frac{2Q}{R^2}\left[\sqrt{z^2+R^2}-z\right].
\end{eqnarray}

To find the potential off the axis of the disk for $r> R$, we first expand the potential on the axis in a binomial expansion in powers of $R/z$.
\begin{eqnarray}
\phi_{\rm disk}(z>R,0^\circ)&=&\frac{2Qz}{R^2}\left[\sqrt{1+R^2/z^2}-1\right]\nonumber\\
&=&\frac{2Q}{R}\sum_1^\infty\left(\begin{array}{c}  1/2\\ n\end{array}\right)
\left(\frac{R}{z}\right)^{2n-1}\nonumber\\
&=&\frac{2Q}{R}\sum_{{\rm even}\,l}^\infty\left(\begin{array}{c}  1/2\\ \frac{l}{2}+1\end{array}\right)
\left(\frac{R}{z}\right)^{l+1}.
\end{eqnarray}

As we did for the ring, we multiply each term in the expansion by $P_l(\cos\theta)$, to get
\begin{eqnarray}
\phi_{\rm disk}(r>R,\theta)&=&2Q\sum_{{\rm even}\,l}^\infty
\left(\begin{array}{c}  1/2\\ \frac{l}{2}+1\end{array}\right)
\left[\frac{R^lP_l(\cos\theta)}{r^{l+1}}\right]
\label{dd}
\end{eqnarray}
for the potential off the axis of the disk, but only in the region $r>R$.

In order find the potential off the axis of a disk for the region $r<R$,
we use the off-axis potential for a uniformly charged ring for each region, $r>R$ and $r<R$,
as given by equations (6) and (7), respectively.

This gives,
\begin{eqnarray}
\phi_{\rm disk}(r<R,\theta)&=&\frac{2}{R^2}\int_0^r\phi_{\rm ring}(r>x,\theta)xdx
+\frac{2}{R^2}\int_r^R\phi_{\rm ring}(r<x,\theta)xdx\nonumber\\
&=&\frac{2Q}{R^2}\sum_{{\rm even}\,l}^\infty\left(\begin{array}{c}  -1/2\\ \frac{l}{2}\end{array}\right)P_l(\cos\theta)\left[\int_0^r\frac{x^{l+1} dx}{r^{l+1}}
+\int_r^R\frac{r^l dx}{x^l}\right]\nonumber\\
&=&\frac{2Q}{R^2}\sum_{{\rm even}\,l}^\infty\left(\begin{array}{c}  -1/2\\ \frac{l}{2}\end{array}\right)P_l(\cos\theta)\left[\frac{r}{l+2}+\frac{r}{l-1}
-\frac{r^l}{(l-1)R^{(l-1)}}\right]\nonumber\\
&=&\frac{2Qr}{R^2}\sum_{{\rm even}\,l}^\infty
\left(\begin{array}{c}  -1/2\\ \frac{l}{2}\end{array}\right)\frac{(2l+1)P_l(\cos\theta)}{(l+2)(l-1)}
\nonumber\\&&
-\frac{2Q}{R}\sum_{{\rm even}\,l}^\infty\left(\begin{array}{c}  -1/2\\ \frac{l}{2}\end{array}\right)
\frac{r^l P_l(\cos\theta)}{(l-1)R^l}.
\end{eqnarray}

Equations (10) and 11) give the  potential of the uniformly charged disk for $r>R$ and $r<R$ respectively.

\section{Electric field  of a uniformly charged disk }

The radial component of the electric field of the  uniformly charged disk is
\begin{eqnarray}
E_r(r,\theta)&=&\partial_r\phi({\bf r},\theta)\nonumber\\
&=&\sum_{{\rm even}\; l}\left(\begin{array}{c} -1/2\\ l/2
\end{array}\right)\frac{2Q(l+1)R^{l}P_l(\cos\theta)}{r^{l+2}},\quad r>R,\\
&=&\sum_{{\rm even}\; l>0}\left(\begin{array}{c} -1/2\\ l/2
\end{array}\right)\frac{2Qlr^{l-1}P_l(\cos\theta)}{R^{l+1}},\quad r<R.
\end{eqnarray}

The angular component of the electric field of the disk is
\begin{eqnarray}
E_\theta(r,\theta)&=&\frac{1}{r}\partial_\theta \phi({\bf r},\theta)
=-\frac{1}{r}\sin\theta\partial_{\cos\theta}\phi({\bf r},\theta)\\
&=&-2Q\sum_{{\rm even}\,l}^\infty
\left(\begin{array}{c}  1/2\\ \frac{l}{2}+1\end{array}\right)
\left[\frac{R^l\sin\theta\partial_{\cos\theta}P_l(\cos\theta)}{r^{l+2}}\right],\quad r>R\\
E_\theta(r,\theta)&=&-2Q\sum_{{\rm even}\; l>0}\left(\begin{array}{c} -1/2\\ l/2\end{array}\right)
\frac{(2l+1)\sin\theta\partial_{\cos\theta}P_l(\cos\theta)}{(l+2)(l-1)R^2}\nonumber\\
&&+2Q\sum_{{\rm even}\; l>0}\left(\begin{array}{c} -1/2\\ l/2\end{array}\right)
\frac{lr^{l-1}\sin\theta\partial_{\cos\theta}P_l(\cos\theta)}{(l-1)R^{l+1}}\nonumber\\
&=&-2Q\sum_{{\rm even}\; l>0}\left(\begin{array}{c} -1/2\\ l/2\end{array}\right)
\left[\frac{(2l+1)}{(l+2)(l-1)R^2}+\frac{lr^{l-1}}{(l-1)R^{l+1}}\right]
\sin\theta\partial_{\cos\theta}P_l(\cos\theta).\nonumber\\
\end{eqnarray}
%where $P_l^1(\cos\theta$ is the firsr order associated Legendre polynomial in the Condon–Shortley phase %convention.
This derivation of the electric field from the Legendre polynomial expansion of the potential is much simpler 
than the procedure  proposed in \cite{11}.
\section{Conclusion}

We have given Legendre polynomial expansions of the elecrostatic potential and electric field of a uniformly charged disk.
This form is simpler and more accessible to graduate students than previous derivations.


\begin{thebibliography}{9}
\bibitem{1} Cayley A 1874 On the potential of the ellipse and the circe Proc. Lond. Math. Soc. s1-6 38–58
\bibitem{2} Durand E 1953 Électrostatique (Les Distributions) vol (Masson)
\bibitem{3} Duboshin G N 1961 The Theory of Attraction (Fizmatlit)
\bibitem{4} Kondratyev B P 2003 Theory of Potential and Equilibrium Figures (Publishing House RHD) (in
Russian)
\bibitem{5} Kondratyev B P 2007 Potential Theory: New Methods and Problems with Solutions (Mir)
\bibitem{6} Krogh F T, Ng E W and Snyder W V 1982 The gravitational field of a disk Celest. Mech. 26
395–4057 30 225–8
\bibitem{7}Lass H and Blitzer L 1983 The gravitational potential due to uniform disks and rings Celest. Mech.
30 225–8
\bibitem{8} Conway J T 2000 Analytical solutions for the Newtonian gravitational field induced by matter
within axisymmetric boundaries Mon. Not. R. Astron. Soc. 316 540–54
\bibitem{9} Bochko V and Silagadze Z K 2020 On the electrostatic potential and electric field of a uniformly
charged disk Eur. J. Phys. 41 045201
\bibitem{10} Ciftja O and Hysi I 2011 The electrostatic potential of a uniformly charged disk as the source of
novel mathematical identities Appl. Math. Lett. 24 1919–23
\bibitem{11}Sagaydak A E and Silagadze Z K
2025 Electrostatic potential of a uniformly charged disk through Green's theorem
Eur. J. Phys. 46  015203-015213 
\end{thebibliography}
\end{document}